\documentclass{article}

\PassOptionsToPackage{square, numbers, compress}{natbib}
\usepackage[preprint]{neurips_2019}
\usepackage[utf8]{inputenc} 
\usepackage[T1]{fontenc}    
\usepackage{microtype}
\usepackage{graphicx}
\usepackage{subfigure}
\usepackage{booktabs, tabularx}
\usepackage{hyperref}
\usepackage{url}  
\usepackage{amsthm}
\usepackage{bm}
\usepackage{bbm}
\usepackage{amssymb}
\usepackage{amsmath}
\usepackage{amsfonts}
\usepackage{dsfont}
\usepackage{nicefrac}
\usepackage{algorithm}
\usepackage{algorithmic}

\DeclareMathOperator*{\argmax}{argmax}

\newtheorem{theorem}{Theorem}
\newtheorem{definition}{Definition}
\newtheorem{proposition}[theorem]{Proposition}
\newtheorem{remark}[theorem]{Remark}
\newtheorem{lemma}[theorem]{Lemma}

\graphicspath{{../Datasets/Artificial/},{../Datasets/Ymovies/},{../Datasets/MovieLens/}}

\newcommand{\sE}{\mathcal{E}}

\newcommand{\sI}{\mathcal{I}}
\newcommand{\sL}{\mathcal{L}}
\newcommand{\sN}{\mathcal{N}}
\newcommand{\sP}{\mathcal{P}}
\newcommand{\sR}{\mathcal{R}}
\newcommand{\sS}{\mathcal{S}}

\newcommand{\sU}{\mathcal{U}}

\newcommand{\sO}{\mathcal{O}}

\newcommand{\mL}{\mathbf{L}}
\newcommand{\mR}{\mathbf{R}}

\newcommand{\mV}{\mathbf{V}}

\newcommand{\bmu}{\bm \mu}
\newcommand{\bK}{{k}}

\newcommand{\util}{\upsilon}
\newcommand{\indexU}{\kappa}
\title{Risk Aware Ranking for Top-$\bK$ Recommendation}
\author{
    Shameem A Puthiya Parambath\\
    QCRI
    \And
    Nishant Vijayakumar\\
    Akamai
    \And
    Sanjay Chawla\\
    QCRI
}
\begin{document}

\maketitle

\begin{abstract}
    We revisit the top-$k$ personalized ranking problem using game-theoretic models and pose the question: what is the best strategy for recommendation ?
    Before answering the question affirmately with risk-seeking strategy, we show that majority of the currently available recommendation algorithms are either risk-neutral or risk-averse.
    We empirically test our algorithms on two benchmark datasets, and show that risk seeking strategy outperforms peers in the most natural experimental settings.
\end{abstract}

\section{Introduction}
\label{sec:intro}
A personalized recommendation agent recommends a ranked list of fixed number of items or objects to a user from a large but finite number of ground set of objects.
The objects to be ranked can be set of videos, music, or alternatives for investment and each object will be associated with a payoff or a utility value.  
The payoff represents the level of satisfaction or the amount of monetary benefit a user obtains by exploring the recommended object like rating value or monetary returns.
For example, in the online video recommendation task, payoff can be a value from the ordinal rating scale from 1 to 5, 1 being very unsatisfactory and 5 being very satisfactory.
In typical recommendation tasks, the number of recommendations to be made, represented as $\bK$, is usually very limited and it is often subject to budget constraints like space constraints.

We assume that there is a linear preference order among the items i.e given a set of items, each user can order the items in decreasing or increasing order of personal preferences. 
We clearly distinguish between the preferences of a user over the items and the actual payoffs obtained from those items.
The preferences of the user for the items are not the same as the payoff values associated with the items.
The highest preferred item might yield in a lower payoff value whereas the least preferred item might yield a higher payoff value.
For example, a user might highly prefer to watch a movie from his favourite director-actor duos but the actual payoff after watching it might be less than other less preferred movies.
Since the items to be recommended are unobserved and payoff values are unknown, there exists an inherent uncertainty associated with the selection of the items.
The recommender system has to take this uncertainty into consideration when making the recommendations.
To be concrete, in the video recommendation setting, assume that the task is to recommend three videos from a ground set of 100 unobserved videos; this results in more than 90 million choices of ranked lists, each having an inherent uncertainty over the payoff after consumption of the recommendation.
Our problem can be formulated as: what is the best strategy a recommender system can follow to select one choice of ranked list containing three videos with the maximum payoff for a user ?

The state-of-the-art ranking algorithms work by either \emph{(i)} predicting the unobserved ordinal rating values approximately \cite{crammer2002pranking}, \emph{(ii)} predicting the pair-wise preferences between items \cite{freund2003efficient,katz2018nonparametric} or \emph{(iii)} predicting the complete linear order among the items \cite{koyejo2013retargeted,gunasekar2016preference}.
In all the aforementioned approaches, recommendation is carried out by first estimating user-wise preference scores for the items and selecting the top-$k$ items from the linear preference order induced by the preference scores.
Thus the aforementioned approaches fail to take into consider the uncertainty in payoff over the consumption of the recommendation.

Here we study the problem of collaborative ranking for recommendations using the expected utility concept from game theory.
We model the uncertainty in payoff over the selection of a choice containing fixed number of unobserved items using lottery space.
We show that risk seeking strategy yields the best top-$\bK$ recommendation in terms of the commonly used ranking metrics in the natural experimental setup. 
In addition, we also show that state of the art algorithms can be classified as either risk-neutral or risk-averse.
\section{Related Work}
\label{sec:rel_work}
The state-of-the-art collaborative ranking algorithms for recommender systems fall under three categories.
Collaborative rankings work by either \emph{(i)} predicting the unobserved ordinal rating values accurately, called point-wise preference estimation \citep{crammer2002pranking} \emph{(ii)} predicting the pair-wise preferences between the items, called pair-wise preference estimation \citep{freund2003efficient,katz2018nonparametric} or \emph{(iii)} using the list-wise ranking models to estimate the preference scores for the corresponding linear order among the items \citep{koyejo2013retargeted,gunasekar2016preference}.
One can use incomplete SVD based methods for point-wise, pair-wise or list-wise ranking by replacing the traditional sum of squares loss fucntion with the appropriate ranking loss.
In general, point-wise approaches predict the absolute preference score of a user for an item using regression or classification models and the final complete ranking is constructed by sorting the preference scores.
Point-wise approches are based on minimizing the Euclidean distance between the observed and predicted rating values and the final induced user-wise ordering might not be accurate \citep{cremonesi2010performance}.
In \citep{steck2010training}, authors proposed a point-wise ranking algorithm based on matrix completion techniques.
\citep{steck2013evaluation} discussed fundamental differences between the two types of evaluation strategies used in the collaborative ranking literature and, in ranking based approaches, advocated for an evaluation strategy which considers both the observed and unobserved items in the test set.
This observation becomes the basis for the evaluation strategy used in this paper.
\citep{shi2010list} proposed an extension to the matrix factorization model for point-wise ranking by optimizing for cross-entropy based loss function.
\citep{balakrishnan2012collaborative} proposed a two stage pair-wise collaborative ranking algorithm where the user and item features extracted in the first step are combined to form a new training set for the second step where a ranking model is trained by minimizing pair-wise logistic loss.
Given pairwise preference data, \citep{park2015preference} proposed an algorithm for estimating the pairwise preference scores on unobserved items assuming that user pairwise preferences are governed by Bradley-Terry-Luce prefernece model \citep{hunter2004mm}.

\citep{koyejo2013retargeted} proposed a method to transform the matrix factorization model to a list-wise ranking model using isotonic regression.
The key idea is to alternatively fit the point-wise matrix factorization estimate to the observed training rating scores and retarget the scores by searching over the space of all monotonic transformations of the scores.
\citep{gunasekar2016preference} extended the algorithm of \citet{koyejo2013retargeted} by regularizing the nuclear norm of the observed rating matrix.
\citet{song2016blind} proposed a non-parametric neighbourhood-based collaborative filtering algorithm for point-wise ranking prediction.
The core idea of the algorithm is based on finding the neighbouring users and items, and the unobserved preference score is predicted using `local taylor approximation'.
Inspired by the work in \citep{song2016blind}, \citet{katz2018nonparametric} proposed a pair-wise ranking algorithm based on neighbourhood models and studied the statistical consistency of the neighbourhood models for preference completion.
It should also be noted that in pair-wise preference estimation methods, final total ranking is constructed by using rank aggregation methods like Copeland ranking procedure \citep{katz2018nonparametric}.
\citet{katz2018nonparametric} also noticed that the neighbourhood models often result in poor top-$k$ ranking compared to matrix completion based models.
\section{Risk-aware Recommendations}
\label{sec:exp_util}
We are given a set $\sU$ of $m$ users and a set $\sI$ of $n$ items such that each user has preference over the set of items.
Furthermore, we assume that each item $i \in \sI$ results in a payoff which differs for different users. 
We make a clear distinction between preference and payoff.
A highly prefered item might not lead to the highest payoff value.
The payoff can be monetary or non-monetary, and in a generic sense it is a measure of satisfaction of the decision maker by consuming the item whereas the preference can be viewed as the prospect of getting maximal satisfaction upon consumption.
The preferences of each user over the set of items are captured using a preference relation.
A preference relation of a user $u$ over the outcomes from $\sI$ is a binary relation denoted by $\succeq_u$.
For $i,j \in \sI$, the preference relation $i \succeq_u j$ means that user $u$ prefers item $i$ to $j$ or is indifferent between the choices.
The relation can be strict, indifferent or both \citep{maschler2013game}, and we also assume that the relation is complete and rational \citep{maschler2013game}.
Similarly, we represent the payoff values for a user $u$ using a utility function $z_u : \sI \rightarrow \Re$ such that $i \succeq j \Leftrightarrow z_u(i) \geq z_u(j)$.
Without loss of generality, we assume that the utility values are positive ordinal rating values.
In practical settings, very small fraction of rating values for the user, item pair is known.
We represent these observed rating values using a sparse $m \times n$ rating matrix $\mR$. 
For a given user $u$, we use $\sO_u \subset \sI$ and $\sN_u \subset \sI$ to represent the set of items for which rating values are observed and unobserved respectively.
The observed rating values are represented as $\sR_u = \mR(u,\sO_u)$.
The pair of observed items and corresponding rating values for user $u$ is denoted as $\sE_u$ i.e. $\sE_u= \{(\indexU_j,\util_j)\}_{j=1\cdots|\sO_u|}$ where $\indexU_j \in \sO_u$ is the $j^{th}$ observed item and $\util_j \in \sR_u$ is the corresponding rating value.
The vector of average observed rating for the items in  $\sI$ is denoted using $\bmu$ i.e. ${\displaystyle \bmu_i = \nicefrac{\sum_{u \in \sU}\mR(u,i)}{\sum_{u \in \sU}[[\mR(u,i)]]}}$ where $[[x]] =  1$ if $x \neq 0$.
Finally, we use calligraphic letters to represent sets and bold capital letters to represent matrices.

\subsection{Top-$\bK$ Recommendation Under Uncertainty}
For each user, recommender systems aim to select and rank top-$\bK$ items from the set of unobserved items.
Since the utility values of the unobserved items are not known, there exist uncertainty over the choice of $\bK$ items outcome.
In game theory, lottery spaces are used to model the uncertain prospect of the outcomes.
A lottery defines probability distribution on $\sI$ with finite support.
\begin{definition}{$\bK$-lottery space:}
    We define \emph{$\bK$-lottery space} as the lottery space corresponds to a choice of exactly $\bK$ items outcome.
\end{definition}
\begin{definition}{ranked $\bK$-lottery space:}
    We define \emph{ranked $\bK$-lottery space} as a $\bK$-lottery space such that $i^{th}$ entry represents the probability that an outcome to be ranked at $i^{th}$ position.
\end{definition}
\begin{proposition}
    For every \emph{ranked $\bK$-lottery} there is an equivalent \emph{${\bK}^2$-lottery}
\end{proposition}
\emph{ranked $\bK$-lottery space} is a convenient way to model the uncertain prospect of selecting choices of top-$\bK$ observed items from the ground set of $n$ items.
To make the concept concrete, in the top-3 video recommendation example given in earlier, \emph{ranked 3-lottery space} consists of probability distribution of all possible choices of three movie combinations from the set of 100 movies i.e. points in a 9-dimensional probability simplex.
The lottery $\mL = [j_p:\nicefrac{1}{3},j_q:\nicefrac{2}{3}, j_r:\nicefrac{1}{6}]$ from a \emph{ranked 3-lottery space} indicates that item $j_p$ is granted with probability $\nicefrac{1}{3}$ to appear in $1^{st}$ position, item $j_q$ is granted with probability $\nicefrac{2}{3}$ to appear at $2^{nd}$ position and item $j_r$ is granted with probability $\nicefrac{1}{3}$ to appear in $3^{rd}$ position.
The items set $(j_p,j_q,j_r)$ is the choice of items associated with lottery $\mL$
We denote the space of \emph{ranked $\bK$-lottery} using $\sL$.

Unlike in the typical game settings, in recommender systems, preferences over the unobserved items or lotteries are not known in advance.
One of the key questions to be answered here is: how can one estimate a ranked $\bK$-lottery for a user over the unobserved item set using the information about the already observed items such that the associated preference relation is rational ?
Here we discuss such a construction of a ranked $\bK$-lottery using Latent Variable models.

\paragraph{Lottery Space in Latent Variable Models}
Latent matrix factorization based models are very popular for feature extraction and a ranked $\bK$-lottery can be estimated based on the similarity between the unobserved items and observed items.
Let $\mV_i \in \Re^d$ be the $d$-dimensional feature vector for the item $i$, one can define a ranked $\bK$-lottery $\mL^u$ for user $u$ as the vector $(p_{\sS_1}^u,p_{\sS_2}^u,\cdots,p_{\sS_k}^u) \in \Re^k$ with
\begin{gather}
    p_{\sS_i}^u = \frac{w(i,\sO_u)}{\sum_{j= 1\dots \bK} w(j,\sO_u)}, \quad \text{where}\nonumber \quad w(i,\sO) = \sum\limits_{l=1}^i\frac{1}{(\bK-l+1)}\sum\limits_{j=1}^{|\sO|}f(\mV_i,\mV_j)
\end{gather}
Here $\sS_i$ is the $i^{th}$ item in the choice $\sS$ of $\bK$ items associated with the lottery $\mL^u$ and $f$ is any non-negative similarity function.
Two possible and commonly used candidates for $f$ are triganometric $\cos$ function and RBF kernel.
The preference relation \(\succeq\) between two lotteries $\mL^u_i, \mL^u_j \in \sL$ is defined as $\mL^u_i \succeq \mL^u_j$ if user $u$ prefers the lottery $\mL_i$ over $\mL_j$.
For brewity, we omit the user superscript hereafter.
\begin{remark}[Learning-To-Rank (l2r) Preference Models]
One can make use of the existing Learning-to-Rank based approaches to estimate the preference scores.
Once the preference scores are estimated, one can perform simple monotonic transformations to convert the preference scores to probabilities over the \emph{ranked $\bK$ lottery space}.
\end{remark}
\subsection{Adaptive Utility Over A Choice}
Similarly, we define the utility of a choice for a user based on three assumptions: \emph{(i)} individual utility of an unobserved item in the choice depends on the payoff values for the similar observed items \emph{(ii)} individual utility of an unobserved item in the choice depends on the payoff values for the same item by other users and \emph{(iii)} utility of an item is recommendation dependent i.e. utility of an item depends on utilities of other items in the choice.
The first hypothesis is the crux of the popular item-item collaborative filtering \cite{sarwar2001item} whereas the second hypothesis is the basis for the user-user collaborative filtering \cite{breese1998empirical} and third hypothesis is a common observation made in diverse and group recommendation tasks \cite{puthiya2016coverage,puthiya2018saga} .
Based on the above assumptions, we propose a utilty function $Z$ as a set function of the payoff values of already observed items, oserved payoff values for the items by other users and the similarity of items to already observed items.
Formally, given the pair of observed items and payoff values $\sE$ and set of unobserved items $\sS$ for a user, utility $Z$ for the set $\sS$ is defined as
\begin{equation}
    Z(\sE,\sS) = \sum_{(\indexU,\util) \in \sE} \util \, g(w(\indexU,\sS)) + \bmu_{\indexU}
    \label{eq:opt_util}
\end{equation}
In the above $\bmu_{\indexU}$ is the average observed payoff for the item $\indexU$ and $g$ is a monotonic non-negative function which we call the risk indicator function.
Similarity based utility functions are very popular in recommender systems \cite{puthiya2016coverage,puthiya2018saga} and the intuition is that the user prefers items similar to the items he/she preferred in the past.
It should be noted that like in the latent variable lottery the above utility function is constructed adaptively.
Using the above utility function, we can base our problem of choice under uncertainty with unknown payoff in terms of uncertainty with known payoff values.

A fundamental theorem from game theory \cite{von1947theory} says that if the preference relation is rational, choosing the best lottery according to the preference relation  amounts to choosing the lottery with the highest expected utility.
We rephrase the theorem by \citet{von1947theory} below
\begin{theorem}[\citep{von1947theory}]
    A rational preference relation \(\succeq\) on \(\sP\) has a corresponding Neumann-Morgenstern utility function and the the best lottery according to the preference relation  amounts to choosing the lottery with the highest expected utility.
\label{th:nm_util}
\end{theorem}

\begin{theorem}
The utility function defined in \ref{eq:opt_util} is a Neumann-Morgenstern utility function
\end{theorem}

Theorem \ref{th:nm_util} suggests that, given a fixed risk measuring function $g$, the top-$k$ recommendation problem with highest expected utility can be framed as the below optimization problem
\begin{equation}
    \max_{\substack{\sS \subseteq \sN \\ \lvert \sS\lvert \leq \bK}} \sum_{i=1}^{|\sS|} p_{\sS_i} Z(\sE,\sS_{-i})
    \label{eq:opt_set}
\end{equation}
where $\sS_{-i}$ is the first $i$ elements of the set $\sS$ and $p_{\sS_i}$ is the $i^{th}$ element of $p_{\sS}$.

\subsection{Risk Based Utility Functions}
In game theory, an agent's strategy can be characterized based on the risk associated with the strategy.
For example in First-Price auctions, an overbidding strategy relative to the Nash Equilibrium bids decreases the expected utility but decreases the risk of losing the bid \cite{holt2002risk}.
Here we propose a risk aware strategy for utility by choosing the risk indicator function to be a risk aware function.

Consider a simple gambling game with random payoff value $\delta$ where one is rewarded with a payoff of $\delta_1$ with probability $0 \leq p \leq 1$ and a payoff of $\delta_2$ with probability $1-p$.
A risk-averse strategy rejects any fair gamble in return of a fixed payoff value which is less than the expected payoff associated with the gambling game whereas a risk-seeking strategy rejects any fixed payoff value in favor of the gamble.
A risk-neutral strategy is indifferent towards the expected payoff of the gambling or the fixed payoff i.e in a risk-neutral strategy, expected payoff is same as the fixed payoff.

Formally, as per the definition of the risk-aversion, the utility function for a risk-averse strategy should satisfy that the expected value of the utility function has to be less than or equal to the fixed payoff value.
Let $h:\Re \rightarrow \Re$ be the real valued utility function; then to be risk-averse $h(\mathbb{E}(x)) \geq \mathbb{E}(h(x))$ should hold.  
Jensen's inequality for concave functions gurantees the above property and thus if we select the utility function to be concave, the user will exhibit risk-averse behaviour.
Similarly, the utility function for a risk-seeking strategy should satisfy that $h(\mathbb{E}(x)) \leq \mathbb{E}(h(x))$.  
Jensen's inequality for convex functions guarantees the above property and thus the payoff function for risk-seeking strategy should be convex.
Finally, the utility function for a risk-neutral strategy should be linear, as the expected value of the payoff function should be equal to the fixed payoff.

\begin{lemma}
    Point-wise, pair-wise and list-wise preference learning algorithms are equivalent to risk-neutral strategy with a unit or constant payoff values.
\end{lemma}

\subsubsection{Exponential Risk Indicator Functions}
\begin{center}
\begin{tabular}{@{}c@{}c@{}}
    \begin{minipage}{0.5\textwidth}
            \centering
            \includegraphics[trim={0pt 18pt 0pt 8pt},clip,width=0.95\textwidth]{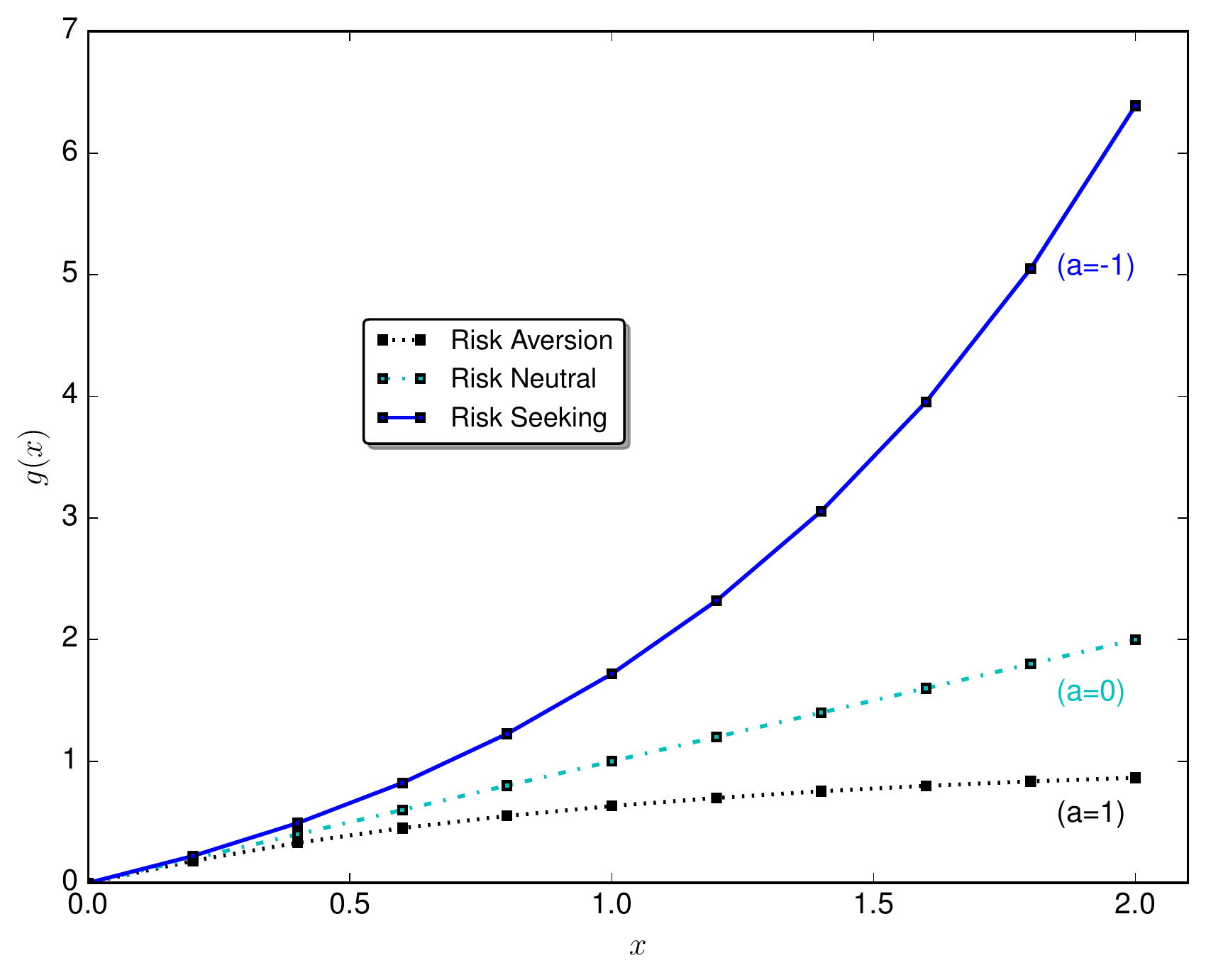}
            \label{fig:risk_plot}
    \end{minipage}
    \begin{minipage}{0.5\textwidth}
        \begin{algorithm}[H]
            \caption{3R Greedy Algorithm}
            \label{alg:greedy}
            \begin{algorithmic}
                \STATE {\bfseries Input:} $\sI$, $\sE_u$, $\bmu$, $\mV$, $f$
                \STATE {$\sI_u = \sI \setminus \sO_u, \sS = \emptyset$}
                \REPEAT
                \STATE {$i^* = \argmax_{i \in \sI_u} \langle p^u_\sS,Z(\sE_u,\sS\cup \{i\})\rangle$}
                \STATE {$\sS = \sS \cup \{i^*\}$}
                \STATE {$\sI_u = \sI_u \setminus \{i^*\}$}
                \UNTIL {$\lvert \sS \lvert = k$}
                \STATE {\bfseries Output:} set of items $\sS$
            \end{algorithmic}
        \end{algorithm}
    \end{minipage}
\end{tabular}
\end{center}
In economics, classes of hyperbolic absolute risk aversion isoelastic functions \cite{ingersoll1987theory} are often used to model the risk behaviour of strategies \cite{ingersoll1987theory}.
Here we use the exponential risk function, a special case of hyperbolic absolute risk functions, as the risk indicator function.
Formally, exponential risk indicator function is defined as
\begin{equation}
    g(x) = 
    \begin{cases}
        \frac{1-e^{-ax}}{a} & \text{if $a \neq 0$} \\
        x & \text{if $a = 0$}
    \end{cases}
    \text{ where $a$ is a constant}
    \label{eq:exp_risk}
\end{equation}
Exponential risk indicator function is monotone and for positive values of $a$, $g(x)$ becomes a monotonic concave function and thus results in a risk-aversion strategy, for $a=0\,, g(x)$ exhibits risk neutral behaviour and for negative values of $a\,, g(x)$ becomes a monotonic convex function and the recommendation exhibits risk seeking strategy.
The plots for exponential risk indicator function for three different settings of $a$ is shown in Figure~\ref{fig:risk_plot}.
Depending on the concave, linear and convex functional forms of the risk indicator function, the optimization problem \ref{eq:opt_set} reduces to submodular function
maximization, modular function maximization or submodular function minimization respectively.
\begin{proposition}
    For a risk-averse risk indicator function, $Z$ is a submodular function and \eqref{eq:opt_set} becomes submodular maximization with uniform matroid constraints
\end{proposition}
\begin{proposition}
    For a risk-neutal risk indicator function, $Z$ is a modular function and \eqref{eq:opt_set} becomes modular function approximation with uniform matroid constraints
\end{proposition}
\begin{proposition}
    For a risk-seeking risk indicator function, $Z$ is a supermodular function and \eqref{eq:opt_set} becomes modular function approximation with uniform matroid constraints
\end{proposition}
It can be easily verified that in all the above three formulations with exponential risk indicator function defined in \eqref{eq:exp_risk}, $Z$ is a monotonic non-decreasing set function.
Also, without loss of generality we can assume that $Z$ is normalized i.e. $Z(\emptyset) = 0$.
The optimization problem in \eqref{eq:opt_set} can be solved efficiently for risk-neutral and risk-aversion strategy by using simple greedy heuristic.
In fact, the very famous result due to \cite{nemhauser1978analysis} states that in case of risk-neutral strategy, greedy heuristic gives the optimal solution and for risk-aversion strategy, the greedy heuristic gives a constant approximation factor, where the approximation factor is equal to $1-\frac{1}{e}$.
The risk-averse strategy has been used in recommendation settings to diversify personal recommendations \cite{puthiya2016coverage} and as aggregation strategy in group recommendations \cite{puthiya2018saga}.

On the other hand, the submodular minimization problem with cardinality constraint is NP-Hard and recent studies have shown that even in the commonly used value oracle model, the problem of minimizing a submodular function under even simpler constraints do not even admit constant or logarithmic approximation factors in polynomial time \cite{svitkina2011submodular,balcan2018submodular,nagano2011size}.
\citet{svitkina2011submodular} proposed a randomized approximation algorithm by randomly sampling subsets of the ground set of items.
If the sampled subset has a large overlap with the optimal set, then the solution is close to the optimal solution with high probability.
But in practical settings such an algorithm can be very expensive.
\citet{nagano2011size} gave a polynomial time algorithm but finds the solution only for a subset of hyperparameter ($k$ in our case), but unfortunately the hyperparameter cannot be specified in advance.
Here we use a simple greedy heuristic to find an approximate solution.
Our experimental results show that such a simple heuristic works well in practice.
The greedy algorithm is given in Algorithm~\ref{alg:greedy}.

\section{Experiments}
\label{sec:exp}
We evaluate our model on two collaborative ranking estimation tasks. 
We use Movielens and Yahoo! Movies dataset for the tasks.
In the Movielens dataset number of users is more than the number of items ($m > n$) whereas in the Yahoo! Movies dataset items outnumber the users ($m < n$).
We show that in both settings, our algorithm outperforms the state-of-the-art algorithms.
Movielens dataset contains 1000209 ratings for 6040 users and 3706 movies and Yahoo! Movies dataset consists of 138310 rating values over 3429 users and 8067 movies. 
In both datasets the ratings are blockwise ordered-taking one of 5 values in the set \{1,2,3,4,5\}.
\begin{table}
    \setlength{\tabcolsep}{2.9pt}
    \begin{tabularx}{\textwidth}{@{}*{2}{>{\centering\arraybackslash}X}@{}}
    \caption{Movielens Top-5 recommendations}
    \label{tab:res1}
    \begin{tabular}{l*{6}{c}}
    \toprule
            Baseline &   SMC &  RMC  &  PMF  &  MOD  &  SCA  & RSR  \\
            \midrule
            NDCG     & 0.864 & 0.877 & 0.933 & 0.931 & 0.930 & 0.935 \\
            MAP      & 0.330 & 0.337 & 0.685 & 0.681 & 0.680 & 0.684 \\
            \bottomrule
        \end{tabular}
        &
    \caption{Yahoo! Top-5 recommendations}
    \label{tab:res2}
    \begin{tabular}{*{6}{c}}
    \toprule
        SMC  &  RMC  &  PMF  &  MOD  &  SCA  &  RSR \\
        \midrule
        0.913 & 0.927 & 0.966 & 0.963 & 0.963 & 0.973 \\
        0.431 & 0.436 & 0.785 & 0.756 & 0.755 & 0.786 \\
        \bottomrule
    \end{tabular}
    \end{tabularx}
    \vspace{-3ex}
\end{table}

\subsection{Evaluation Methodologies}
Following \citep{steck2013evaluation}, we use two types of evaluation methodologies for recommender systems.
In the first methodology, only the observed items in the test set are considered for evaluation whereas in the second methodology the entire test set is used for evaluation.
An in-depth analysis of the two evaluation strategies has been given in \citet{steck2013evaluation} and authors advocated that first methodology is suitable when recommendation is considered as a rating prediction problem and the second methodology is suitable when recommendation is considered as a ranking problem.
Publically available l2r models for collaborative filtering \citep{koyejo2013retargeted,gunasekar2016preference,weimer2008cofi} use the first evaluation methodoloy.
Hence, we evaluate our model using both evaluation strategies and show that our method improves over the baselines when recommendation is viewed as a ranking problem. 

In both evaluation strategies, we followed the experimental protocol employed in recommender systems research \citep{cremonesi2010performance,puthiya2018saga} and also used by our baseline algorithms \citep{gunasekar2016preference,puthiya2016coverage}.
We split the data into training and test set such that 5\% of the original data goes into testing and the remaining goes into training.
The split is carried out five times in a manner that both the training and test set span the entire user and movie set in each split.
The reported results are the average over the five splits.
The user and item features are extracted from the observed rating matrix using matrix factorization.
Following \citep{steck2013evaluation}, we used regularized weigted non-negative matrix factorization to extract the user and item features.
We used the Frobenius norm of the factor matrices as a lower bound of the trace norm of the rating matrix to regularize the matrix factorization objective \cite{srebro2005maximum}.

\subsection{Evaluation Metrics}
The performance of the ranking tasks are evaluated on three metrics: two ranking metrics and one coverage metric.
We use Normalized Discounted Cumulative Gain (NDCG) and Mean Average Precision (MAP) as the ranking metrics and Topic Coverage as the coverage metric.
Both NDCG and MAP evaluate the correcteness of the proposed ranking in the top K of the recommendation.
We use the raw observed rating values in the test set to calculate the NDCG.
MAP is a binary ranking metric and we discretized the observed rating value to calculate the MAP value for the test set.
We used binary discretization such that rating values of 4 and 5 are deemed as relevant and as irrelevant otherwise.
Topic Coverage is a commonly employed evaluation metric in the diverse recommendation tasks.
To calculate Topic Coverage, we used the genres associated with the movies.
Topic Coverage is defined as the ratio of the number of relevant genres (generes associated with the relevant movies) covered in the top k ranking and the number of relevant genres covered by the observed test items.

\subsection{Baselines}
In addition to the risk-neutral (MOD) and risk-averse (SUB) versions of the proposed algorithm, we used three Learning to Rank based models.
Learning to Rank based baselines include two list-wise Learning to Rank algorithms: Retargeted matrix factorization for collaborative filtering (SMC) \cite{koyejo2013retargeted} and Nuclear Norm Regularized Retargeted Matrix Factorization for Collaborative Filtering (SMC) \cite{gunasekar2016preference} and a point-wise Learning to rank algorithm (PMF) based on weighted matrix factorization \cite{steck2010training}.
For RMC and SMC, we used publically available code with default parameter settings. 
Other l2r baslines like Cofi-Rank \citep{weimer2008cofi} is exluded as it performed suboptimally compared to our other l2r baselines on same experimental settings \citep{gunasekar2016preference,puthiya2016coverage}.
We use Rank aware Risk seeking Recommendation (3R) to denote the proposed algorithm.

\subsection{Results \& Discussion}
\begin{figure}
    \label{fig:results}
        \begin{tabular}{cc}
            \includegraphics[trim={0pt 5pt 0pt 5pt},clip,width=0.5\textwidth]{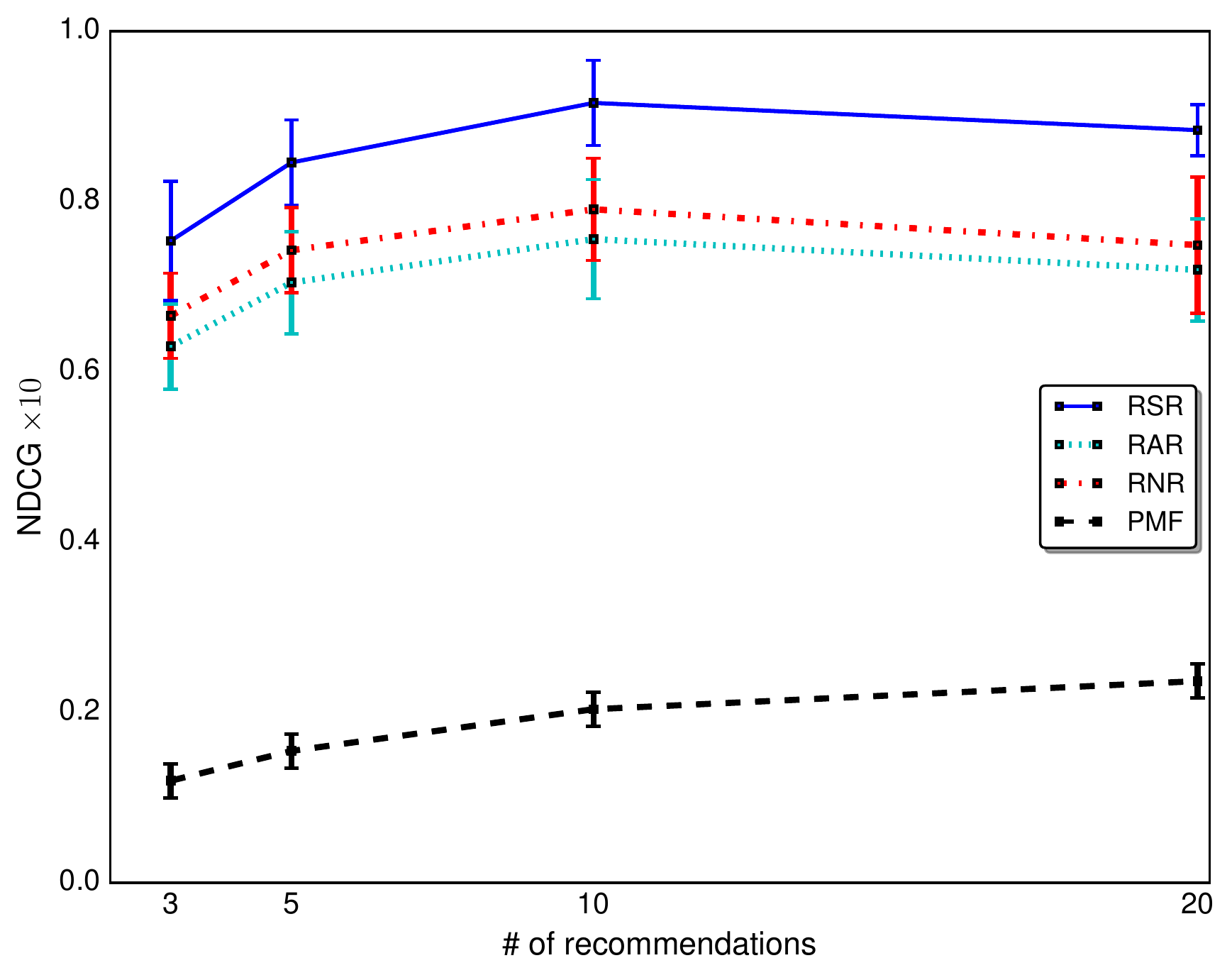} & \includegraphics[trim={0pt 5pt 0pt 5pt},clip,width=0.5\textwidth]{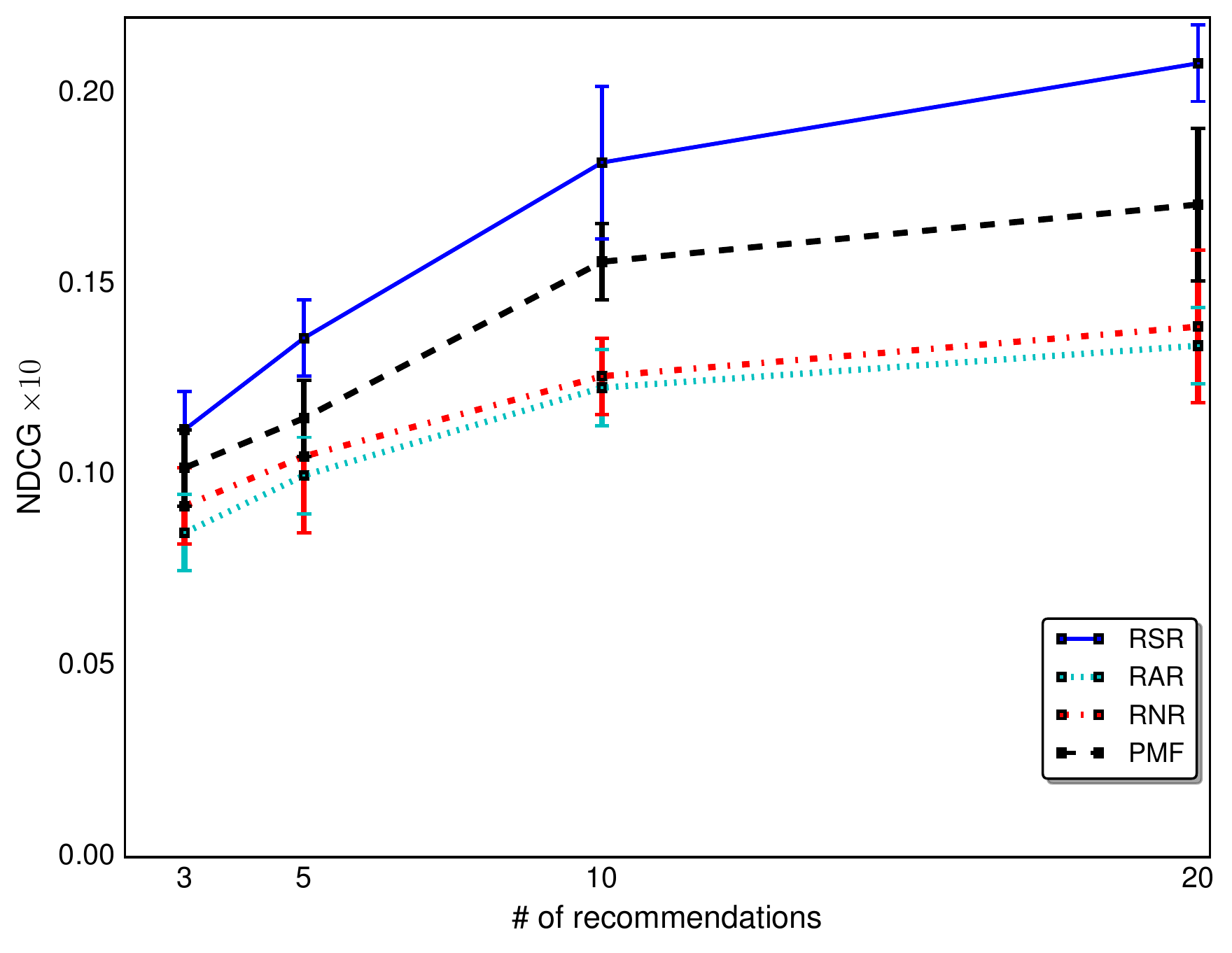}\\
            \includegraphics[trim={0pt 5pt 0pt 5pt},clip,width=0.5\textwidth]{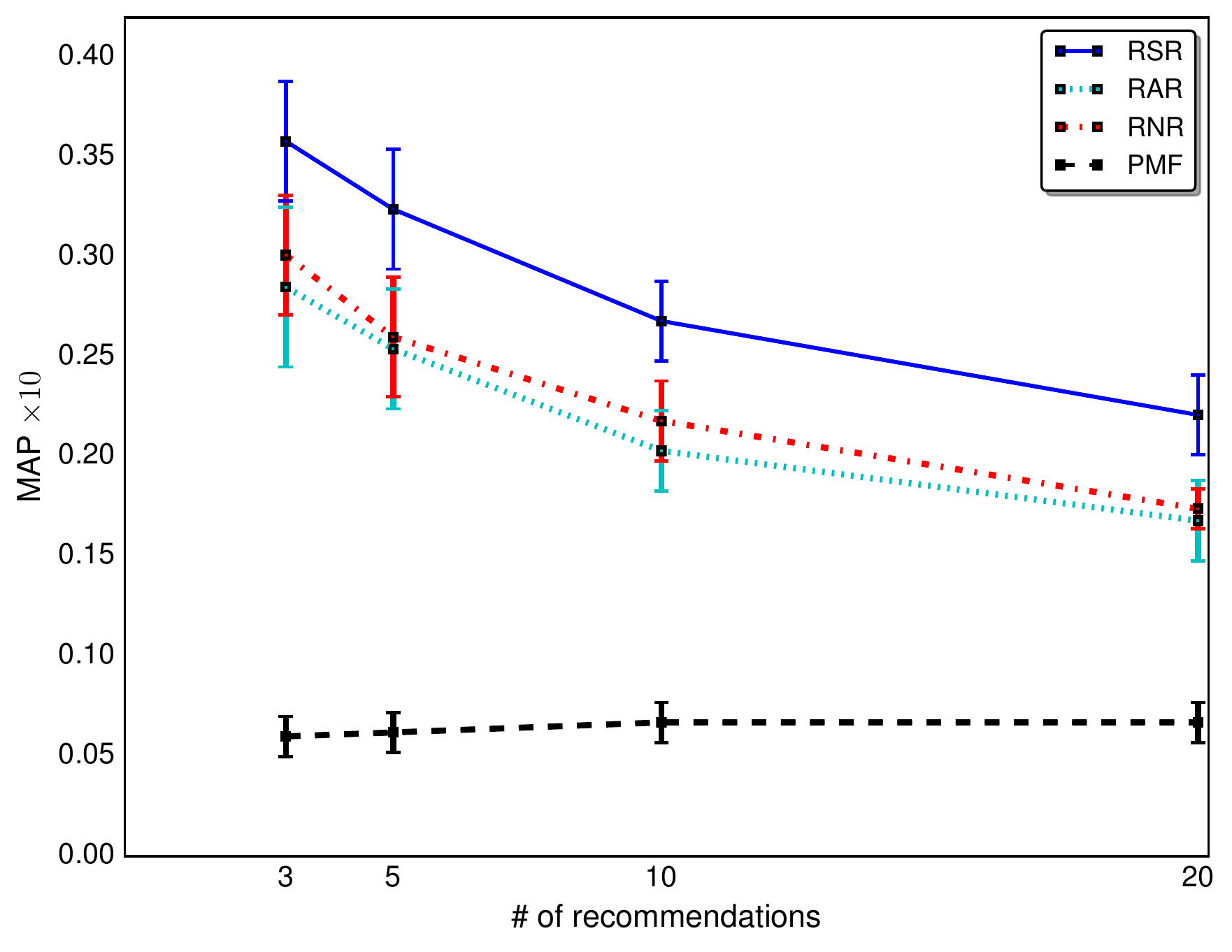}  & \includegraphics[trim={0pt 5pt 0pt 5pt},clip,width=0.5\textwidth]{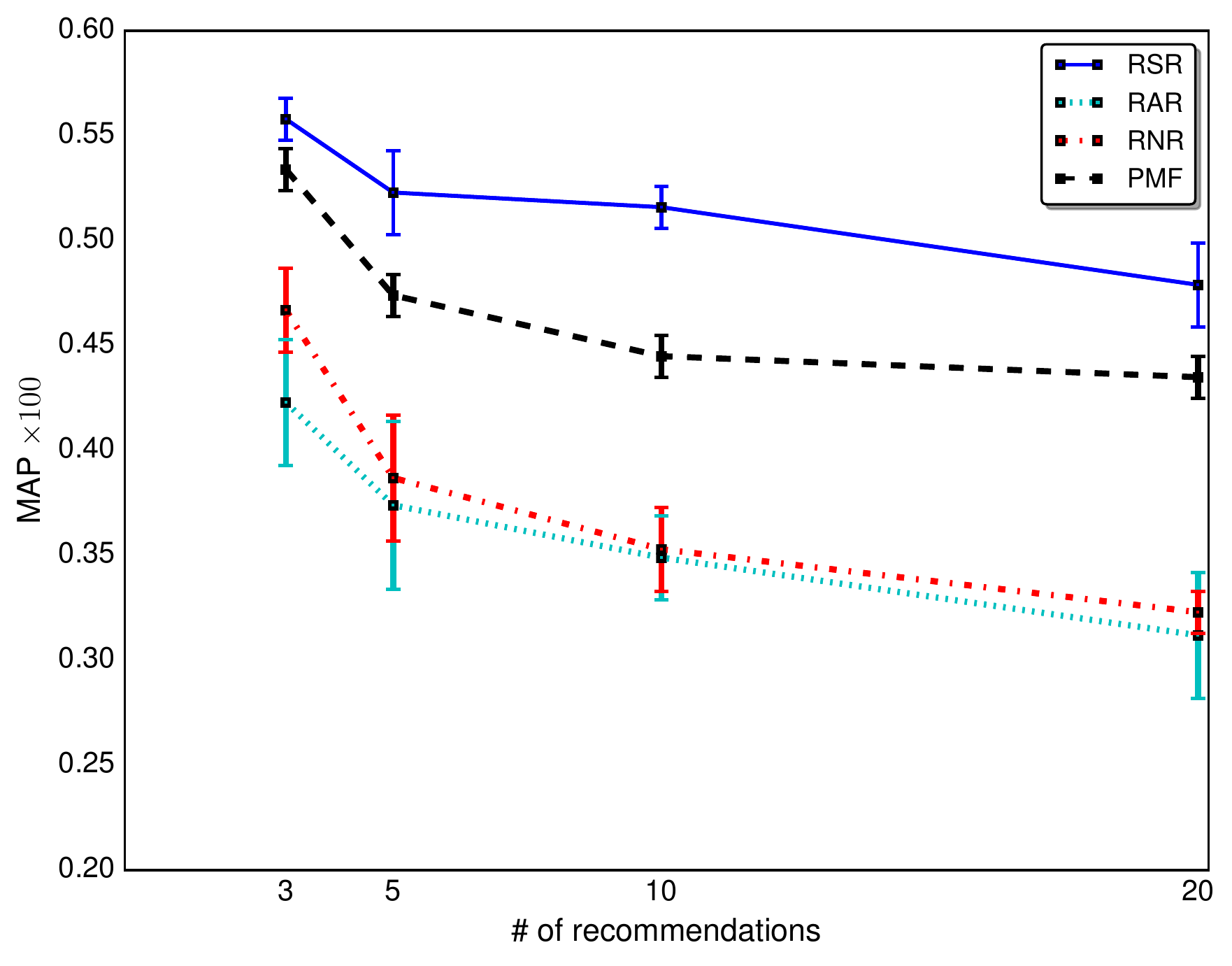} \\
            \includegraphics[trim={0pt 5pt 0pt 5pt},clip,width=0.5\textwidth]{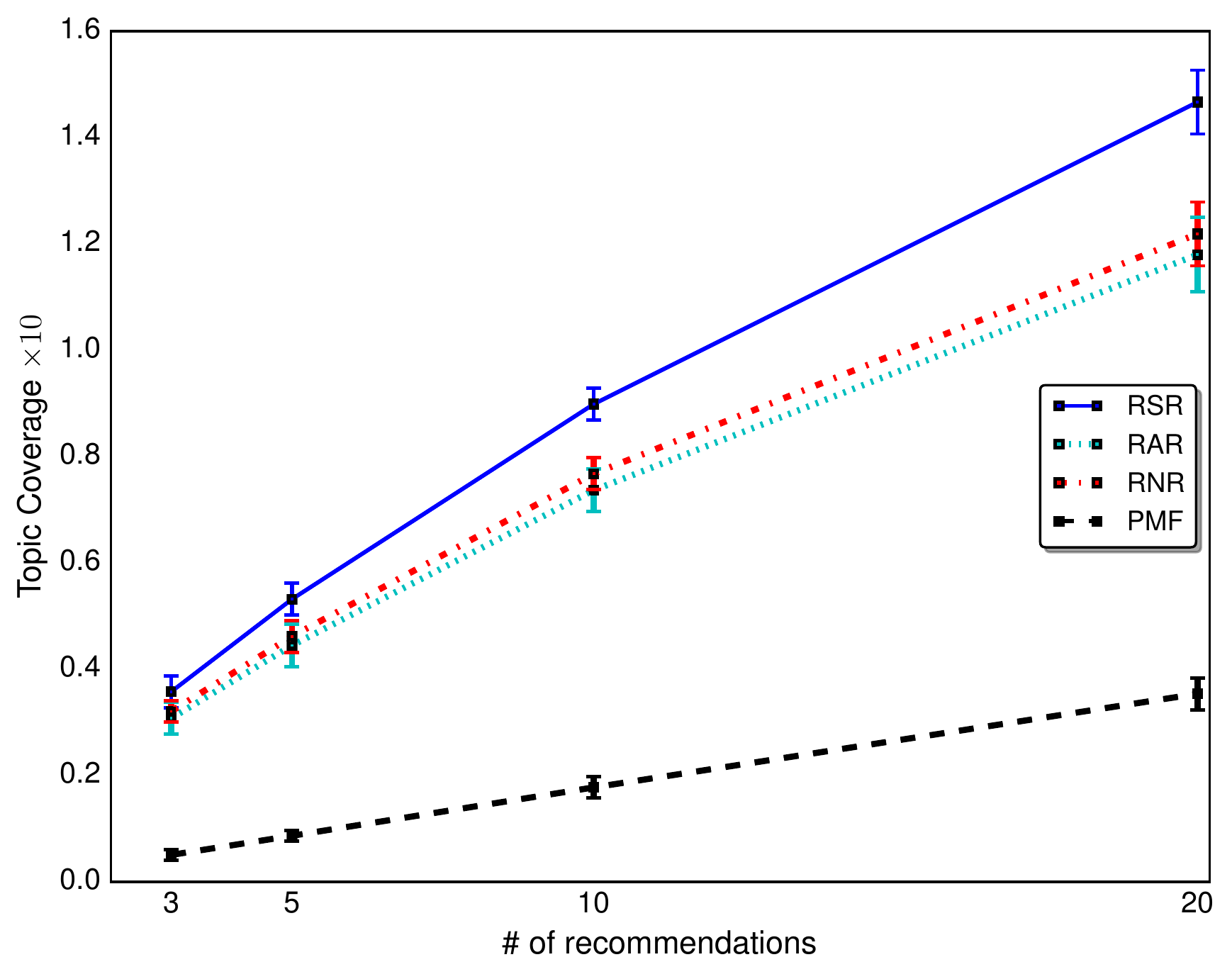}   & \includegraphics[trim={0pt 5pt 0pt 5pt},clip,width=0.5\textwidth]{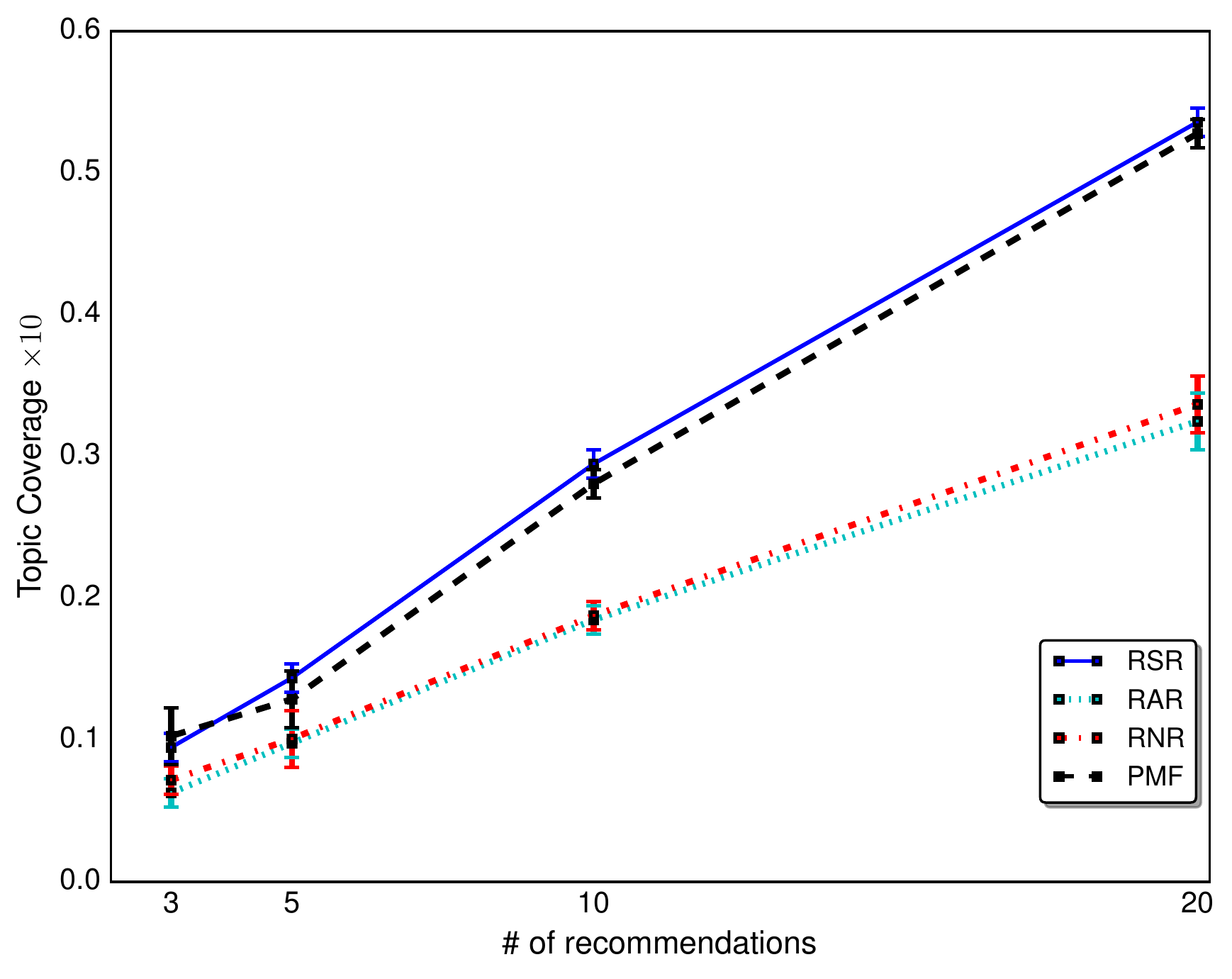}
        \end{tabular}
    \caption{Ranking Performance on Movielens and Yahoo! Movies. Left column compares the performance of different algorithms on the Movielens data whereas the right column compares the performance on the Yahoo! Movies. \textbf{3R} represents the proposed algorithm}
\end{figure}
\hspace{\parindent} \textbf{Recommendation As Rating Prediction}
Experimental results for top-5 recommendations based on the first evaluation strategy on Movielens and Yahoo! Movies are summarized in Table~\ref{tab:res1} and Table~\ref{tab:res2} respectively.
Though this evaluation strategy is better suited for rating prediction tasks than ranking tasks, our proposed algorithm performs as good as the competitors. 
Surprisingly, SMC and RMC perform marginally lower than simple weighted matrix factorization.
The same trend is noticed for different values of $k$ (results are in supplementary file).

\hspace{\parindent} \textbf{Recommendation As Ranking Prediction}
Here we report the results based on the second and the most natural evaluation strategy of considering recommendation as top-$\bK$ ranking problem.
Figure~\ref{fig:results} presents the results of our evaluation on the MovieLens (left column) and the Yahoo! Movies (right column) with the corresponding standard deviation bars.
Unfortunately, SMC and RMC is hard-coded to evaluate only on the observed test items, hence we did not report the results here but it can be concluded from the results of the first evaulation strategy that it performs poorly.
As expected, the proposed risk seeking strategy outperforms the state-of-the-art collaborative ranking algorithms.
We get a performance improvement of more than 10\% in terms of NDCG values.
In case of Movielens, as the number of recommendations are increased from 3 to 10, the NDCG values increased proportionally but the NDCG value for the top 20 recommendation is marginally lower than top 10 recommendation.
This trend can be noticed for all the baselines we used.
The average number of users with more than 10 ratings in the test set is 1353, and the remaining 4687 users did not contribute to NDCG values for the top 20 recommendations, and subsequently NDCG values for top 20 recommendation is less than the top 10 recommendations.
The same trend is not visible in case of Yahoo! Movies.
In case of Yahoo! Movies NDCG values for top 20 recommendations are higher than top 10 recommendations though the average number of users with more than 10 ratings in the test set is low.
This can be explained using the sparsity of the Yahoo! Movies dataset.
Yahoo! Movies dataset is 99.50\% sparse whereas Movielens dataset is 95.53\% sparse and the average number of users with more than 3 ratings is less than 750 in the test set.
Also, for smaller values of the number of recommendations, the performance gain using the proposed risk seeking strategy is marginal and as the number of recommendation size increases the results becomes statistically significant.

It can be noted that as the recommendation size increases MAP values for both the Movielens and Yahoo! Movies decreases.
This is due to the low discretization threshold we use for the binary ranking metric MAP. 
We deem an item as relevant only when the observed rating value is 4 or 5.
Though the average number of relevant ratings per user in Movielens is 21 and Yahoo! Movies is 7, the median relevant ratings per user is 4 and 2 respectively for the Movielens and Yahoo! Movies dataset.
Hence as the recommendation size increases, the MAP values decreases on average.

\section{Conclusion}
\label{sec:conc}
We presented an algorithm for top-$\bK$ ranking for recommendation based on the concepts of expected utility maximization in game theory.
Our algorithm is based on employing risk seeking utility function.
Experiments on benchmark datasets showed that the algorithm performs well, compared to strong baselines.
\bibliographystyle{icml2019}
\bibliography{paper}
\end{document}